\documentclass{article}

\usepackage{amsmath,amssymb,amsthm}
\usepackage[dvips]{graphicx}

\ifx\mathmr\undefined\let\mathmr=\mathrm\fi
\ifx\mathmb\undefined\let\mathmb=\mathbf\fi

\newcommand{\reals}{\mathbb{R}}% real line
\newcommand{\naturals}{\mathbb{N}}% natural numbers
\newcommand{\sphere}{\mathbb{S}}% the sphere/circle
\newcommand{\E}{\mathcal{E}}% set E
\newcommand{\G}{\mathcal{G}}% set G
\newcommand{\K}{\mathcal{K}}% set K
\newcommand{\N}{\mathcal{N}}% set N
\newcommand{\T}{\mathcal{T}}% set T
\newcommand{\U}{\mathcal{U}}% set U
\newcommand{\V}{\mathcal{V}}% set V
\newcommand{\W}{\mathcal{W}}% set W
\renewcommand{\L}{\mathcal{L}}% Lorentz group (overriding \L char)
\renewcommand{\l}{\mathfrak{l}}% Lie algebra (overriding \l char)
\newcommand{\cov}[1]{\nabla_{#1}}% covariant differentiation
\newcommand{\id}{\mathmr{d}}% integral d
\renewcommand{\d}{\partial}% partial d (overriding \d accent)
\newcommand{\idd}[1]{\frac{\id}{\id #1}}% total derivative
\newcommand{\dd}[1]{\frac{\d}{\d #1}}% partial derivative

\newcommand{\dbar}{\bar{d}}
\newcommand{\Nbar}{\bar{\N}}
\newcommand{\tbar}{\bar{t}}
\newcommand{\lambdabar}{\bar{\lambda}}
\newcommand{\gammabar}{\bar{\gamma}}
\newcommand{\pbar}{\bar{p}}
\newcommand{\OMEGA}{\boldsymbol{\Omega}}
\newcommand{\OMEGAbar}{\bar{\OMEGA}}
\newcommand{\LAMBDA}{\boldsymbol{\Lambda}}
\newcommand{\Psibar}{\bar{\Psi}}
\newcommand{\phibar}{\bar{\phi}}
\newcommand{\mubar}{\bar{\mu}}
\newcommand{\xibar}{\bar{\xi}}
\newcommand{\dtilde}{\tilde{d}}% almostsemimetric dtilde
\newcommand{\abs}[1]{\lvert#1\rvert}% absolute value
\newcommand{\ABs}[1]{\Bigl\lvert#1\Bigr\rvert}% absolute value, Big
\newcommand{\ABS}[1]{\biggl\lvert#1\biggr\rvert}% absolute value, bigg
\newcommand{\norm}[1]{\lVert{#1}\rVert}% norm
\newcommand{\ver}{\operatorname{ver}}% vertical component
\newcommand{\fc}{\mathmb}% frame component arrays
\newcommand{\canform}{\boldsymbol{\theta}}% canonical 1-form
\newcommand{\connform}{\boldsymbol{\omega}}% connection form
\newcommand{\omicron}{o}

\renewcommand{\b}[1]{b\nobreakdash-\hspace{0pt}#1}% prevent linebreaks in b-xxx

\newcommand{\OMbar}{\hspace{0.18em}\overline{\hspace{-0.18em}OM}}
\newcommand{\dOM}{\partial OM}
\newcommand{\clbM}{\mathrm{cl}_{b}M}
\newcommand{\dbM}{\partial_b M}
\newcommand{\dtop}{\partial}
\newcommand{\cl}{\overline}

\renewcommand{\sb}{\!_}% subscript after superscript

\newtheorem{theorem}{Theorem}[section]
\newtheorem{lemma}[theorem]{Lemma}
\newtheorem{proposition}[theorem]{Proposition}
\newtheorem{corollary}[theorem]{Corollary}
\newtheorem{conjecture}[theorem]{Conjecture}
\newtheorem{definition}[theorem]{Definition}
\newtheorem*{claim}{Claim}

\begin{document}

\title{On Imprisoned Curves and b-length\\ in General Relativity}
\author{Fredrik St{\aa}hl%
	\thanks{%
		\protect\parbox[t]{100mm}{%
			Department of Mathematics, University of Ume{\aa}, 
					S-901 87 Ume{\aa}, Sweden.\newline
			E-mail address: Fredrik.Stahl@math.umu.se}}}
\maketitle

\begin{abstract}
This paper is concerned with two themes: imprisoned curves and the 
\b{length} functional.  In an earlier paper by the author, it was 
claimed that an endless incomplete curve partially imprisoned in a 
compact set admits an endless null geodesic cluster curve.  
Unfortunately, the proof was flawed.  We give an outline of the 
problem and remedy the situation by providing a proof by different 
methods.  Next, we obtain some results concerning the structure of 
\b{length} neighbourhoods, which gives a clue to how the geometry of a 
spacetime $(M,g)$ is encoded in the pseudo-orthonormal frame bundle 
equipped with the \b{metric}.  We also show that a previous result by 
the author, proving total degeneracy of a \b{boundary} fibre in some 
cases, does not apply to imprisoned curves.  Finally, we correct some 
results in the literature linking the \b{lengths} of general curves in 
the frame bundle with the \b{length} of the corresponding horizontal 
curves.
\end{abstract}

\section{Introduction}
\label{sec:intro}

In general relativity, the concept of \emph{\b{length}} (or 
\emph{generalised affine parameter length}) is essential, in that a 
spacetime is said to be singular if it contains a curve that cannot be 
extended to a curve with infinite \b{length}.  This paper has two main 
themes: properties of the \b{length} functional and of imprisoned 
incomplete curves.

We start by giving some preliminary definitions in 
section~\ref{sec:preliminaries}.  After that, we give some comments on 
the variational theory of the \b{length} functional.  In 
\cite{Stahl:variational}, the author stated a theorem linking 
\b{length} extremals to geodesics of $(M,g)$.  However, as pointed out 
by V.~Perlick \cite{Perlick:b-length}, the proof in 
\cite{Stahl:variational} is flawed.  It turns out that \b{length} 
extremals will \emph{not} be geodesics, except in very special cases.  
We give an outline of the argument in section~\ref{sec:imp-variation}.

In section~\ref{sec:cluster}, we study cluster curves of sequences of 
curves with \b{length} tending to~0.  We establish a technical result 
that will be used in section~\ref{sec:nullcone}, which also allows us 
to settle the issue of Theorem~3 in \cite{Stahl:variational}: 
incomplete and endless curves which are partially imprisoned in a 
compact set admit null geodesic cluster curves.  This is in agreement 
with the corresponding result for totally imprisoned curves 
\cite{Schmidt:local-b-completeness}.

We then turn to the study of \b{distance} neighbourhoods in 
section~\ref{sec:nullcone}, the main idea being to find information 
about how the geometry of $(M,g)$ is encoded in the pseudo-orthonormal 
frame bundle $OM$ with \b{metric} $G$.  Since the \b{length} of a null 
geodesic segment can be made arbitrarily small by a suitable boost of 
the initial frame, the \b{neighbourhoods} of a given point contain the 
light cone of that point.  If we restrict attention to compact sets 
without imprisoned null geodesics, the points on the light cone are 
the only ones having this property.  If we allow the set to `touch' 
the \b{boundary}, by allowing it to be open or to contain imprisoned 
curves, the situation is not so clear.  We provide some illustrations 
by means of examples in the Minkowski, Misner and Robertson-Walker 
spacetimes in section~\ref{sec:examples}.

In \cite{Stahl:degeneracy} it was shown that the fibre over a 
\b{boundary} point $p$ is completely degenerate, given that the frame 
components of the curvature and its first derivative along a 
horizontal curve ending at $p$ diverge sufficiently fast.  Since an 
incomplete endless imprisoned curve ends at a \b{boundary} point, one 
might ask if the methods of \cite{Stahl:degeneracy} is applicable to 
that situation.  In section~\ref{sec:imp-fibre} we show that this is 
not the case.

Finally, we give a result on the \b{length} of general curves in the 
pseudo-orthonormal frame bundle $OM$ in relation to the \b{length} of 
horizontal curves in Appendix~\ref{sec:horizontal}.  In the 
literature, it is sometimes stated that the \b{length} of a horizontal 
curve is less than or equal to the \b{length} of general curve, if the 
two curves start at the same point in $OM$ and the projections to $M$ 
coincide \cite{Dodson:edge-geometry, Clarke:analysis-sing}.  We show 
that this is not really the case, but that a similar estimate can be 
established, which is sufficient for the applications in 
\cite{Dodson:edge-geometry} and \cite{Clarke:analysis-sing}.

\section{Preliminaries}
\label{sec:preliminaries}

The basic object in general relativity is spacetime, which is a pair 
$(M,g)$ where $M$ is a smooth 4-dimensional connected orientable and 
Hausdorff manifold and $g$ is a smooth Lorentzian metric on $M$.

We need to define some concepts relating to curves $\gamma:I\to M$.  
Here $I$ is an interval in $\reals$, possibly infinite.  Suppose that 
$\gamma$ is a future directed curve and that $\U\subset M$.  A point 
$p\in\U$ is a \emph{future endpoint} of $\gamma$ if for any 
neighbourhood $\V$ of $p$ in $\U$ there is a parameter value $t_0\in 
I$ such that $\gamma(t)\in\V$ for every $t\in I$ with $t\ge t_0$.  A 
curve without future endpoint in $\U$ is said to be \emph{future 
endless} in $\U$.  We also say that a geodesic $\gamma$ is 
\emph{future inextendible} in $\U$ if $\gamma$ cannot be extended to 
the future as a geodesic in $\U$.  In an open set, a geodesic is 
inextendible if and only if it is endless, while if the set isn't open 
an inextendible geodesic may have endpoints on the boundary.  Of 
course, there are obvious analogues of these definitions with `future' 
replaced by `past'.  We will usually leave out the temporal adjective, 
the direction being defined by the context.

To reduce index clutter we will, somewhat sloppily, denote a 
subsequence by saying that, e.g., $\{x_j\}$ is a subsequence of 
$\{x_i\}$.  We then mean that $j$ takes values in an index set that is 
a subset of the index set of $i$. 

We will deal extensively with sequences of curves $\{\lambda_i\}$.  We 
say that $\{\lambda_i\}$ \emph{converges to a point} if there is a 
point $p\in M$ such that for any neighbourhood $\U$ of $p$, there is 
an $N\in\naturals$ such that $\lambda_i$ is contained in $\U$ for all 
$i>N$.  There is some confusion in the literature concerning the 
terminology used for the various concepts of convergence of a sequence 
of curves.  Here we choose to reserve the term `limit' for the 
stronger type of convergence which is termed `convergence' 
in~\cite{Wald:GR}, and replace `limit' with `cluster', which the 
author feels is more appropriate (see also~\cite{Hawking-Ellis}).  We 
say that $p$ is a \emph{limit point} of a sequence of curves 
$\{\lambda_i\}$ if for every neighbourhood $\U$ of $p$, there is an 
$N\in\naturals$ such that $\lambda_i$ intersects $\U$ for each $i>N$.  
Similarly, we say that $p$ is a \emph{cluster point} of 
$\{\lambda_i\}$ if every neighbourhood $\U$ of $p$ intersects 
infinitely many $\lambda_i$.  Alternatively, a cluster point of 
$\{\lambda_i\}$ is a limit point of some subsequence of 
$\{\lambda_i\}$.  A curve $\gamma$ is said to be a \emph{limit curve} 
of $\{\lambda_i\}$ if all points on $\gamma$ are limit points of 
$\{\lambda_i\}$.  Finally, $\gamma$ is a \emph{cluster curve} of 
$\{\lambda_i\}$ if $\gamma$ is a limit curve of some subsequence of 
$\{\lambda_i\}$.  Note that being a `cluster curve' is a stronger 
restriction than being a `curve of cluster points'.

Next we define what is meant by imprisoned curves.  A curve is said to 
be (past or future) \emph{totally imprisoned} in a compact set $\K$ if 
it is completely contained in $\K$ (to the past or the future), and 
\emph{partially imprisoned} if it intersects $\K$ an infinite number 
of times.  In~\cite{Hawking-Ellis}, these concepts are defined only 
for causal curves, but they can be applied to general curves as well.  
The case of interest is of course when the imprisoned curve is endless 
and incomplete.

To define what is meant by a curve being incomplete, we need to 
define what we mean by the length of a curve. Given a curve 
$\gamma:I\to M$ and a pseudo-orthonormal frame $E_0$ at some point 
of $\gamma$, we define the \emph{\b{length}} or \emph{generalised 
affine parameter length} as 
\begin{equation}\label{def:b-length}
	l(\gamma,E_0) := \int_{I} \abs{\fc{V}} \,\id t,
\end{equation}
where $\abs{\fc{V}}$ is the Euclidian norm of the component vector 
$\fc{V}$ of the tangent vector of $\gamma$ in the frame $E$ resulting 
from parallel propagation of $E_0$ along $\gamma$ 
\cite{Schmidt:b-boundary,Hawking-Ellis}.

Because the \b{length} of a curve is dependent on a parallel frame 
along the curve, it is convenient to introduce the bundle of 
pseudo-orthonormal frames $OM$.  $OM$ is principal fibre bundle over 
$M$ with the Lorentz group $\L$ as its structure group, and we write 
the right action of an element $\fc{L}\in\L$ as $R_{\fc{L}}\colon 
E\mapsto E\fc{L}$ for any $E\in OM$.  Since $OM$ is a principal fibre 
bundle, there is a canonical 1-form $\canform$ on $OM$, taking values 
in $\reals^4$.  Also, the metric on $M$ induces a connection form 
$\connform$ on $OM$ which takes values in the Lie algebra $\l$ of 
$\L$.  Using these two forms we may define a Riemannian metric on 
$OM$, the \emph{Schmidt metric} or \emph{\b{metric}}, by
\begin{equation}\label{def:G}
	G(X,Y) := \langle\canform(X),\canform(Y)\rangle_{\reals^4}
	      	+ \langle\connform(X),\connform(Y)\rangle_{\l},
\end{equation}
where $\langle\cdot,\cdot\rangle_{\reals^4}$ and 
$\langle\cdot,\cdot\rangle_{\l}$ are Euclidian inner products with 
respect to fixed bases in $\reals^4$ and $\l$, respectively.  There is 
still some arbitrariness in the choice of these fixed bases, but it 
can be shown that a change of bases transforms the \b{metric} to a 
uniformly equivalent metric 
\cite{Schmidt:b-boundary,Dodson:edge-geometry}.

We now define the \b{length} of a general curve $\gammabar:I\to OM$ as 
the metric length of $\gammabar$ with respect to $G$.  In other words, 
the \b{length} of\, $\gammabar$ is
\begin{equation}\label{def:b-length-OM}
	l(\gammabar) := \int_I \Bigl( 
										\abs{\canform(\dot{\gammabar})}^2 + 
										\norm{\connform(\dot{\gammabar})}^2
										\Bigr)^{1/2} \,\id t,
\end{equation}
where $\abs\cdot$ and $\norm\cdot$ are Euclidian norms in $\reals^4$ 
and $\l$, respectively, and $\dot{\gammabar}$ denotes the tangent of 
$\gammabar$.  For horizontal curves, \eqref{def:b-length-OM} is in 
agreement with the previous definition \eqref{def:b-length}, in the 
following sense: if $\gammabar$ is the horizontal lift of a curve 
$\gamma$ with parallel frame $E$, then $l(\gammabar)=l(\gamma,E)$.  We 
also write $d(E,F)$ for the \b{metric} distance between two points 
$P,Q\in OM$ and $B_r(P)$ for the open ball in $OM$ with centre at $P$ 
and \b{metric} radius $r$.

The metric $G$ turns $(OM,G)$ into a Riemannian manifold, in 
particular, $OM$ is a metric space with respect to the topological 
metric $d$.  One may therefore construct the Cauchy completion 
$\OMbar$ of $OM$, and we write $\dOM=\OMbar\setminus OM$.  By 
extending the right action of $\L$, it is possible to project $\OMbar$ 
to an extension $\clbM$ of $M$.  The \b{boundary} of $M$ is then 
defined as $\dbM=\clbM\setminus M$.  We refer to 
\cite{Schmidt:b-boundary}, \cite{Dodson:edge-geometry} or 
\cite{Hawking-Ellis} for the details.

Finally, we denote the topological boundary of a set $\U$ by 
$\dtop\U$ and the topological closure by $\cl\U$. If $\U$ a subset of $OM$, 
$\cl\U$ means the usual closure of $\U$ in $OM$ and not in the Cauchy 
completion $\OMbar$, unless stated otherwise.

\section{Imprisoned curves and variations of \protect\b{length}}
\label{sec:imp-variation}

In~\cite{Stahl:variational}, the author studied local variations of 
the \b{length} functional \eqref{def:b-length}, the primary purpose 
being to apply the result to imprisoned curves.  The result was that 
in sufficiently small globally hyperbolic sets, causal curves of 
minimal \b{length} are geodesics.  However, this statement is false, 
and there is an error in the main argument 
of~\cite{Stahl:variational}, as pointed out by V.~Perlick 
\cite{Perlick:b-length}.  We give an outline of the argument here, 
this section being completely due to V.~Perlick.  The author of this 
paper accepts the responsibility for any errors, of course.

For the moment, we disregard the presence of a Lorentz metric $g$ and 
view $M$ as a smooth manifold with smooth connection $\nabla$ and 
without torsion.  Let $p,q\in M$ and fix a frame $E_p$ at $p$.  We 
consider a variational principle where the trial paths are smooth 
curves of the form $\lambda\colon[0,a]\to M$ from $p$ to $q$, and the 
functional to be extremised is the \b{length} $l(\lambda,E_p)$, given by 
\eqref{def:b-length}.

\begin{proposition}\label{pr:b-extremal}
Let $\lambda\colon[0,a]\to M$ be a curve from $p$ to $q$ in $M$.  
Without loss of generality we may assume that $\lambda$ is 
parameterised by \b{length} $t$.  Let $\fc{V}^i$ and 
$\fc{R}^i\sb{jkl}$ be the components of the tangent of $\lambda$ 
and the Riemann tensor, respectively, in the frame $E$ obtained by 
parallel propagation of $E_p$ along $\lambda$.  Then $\lambda$ is an 
extremal of the \b{length} functional only if
\begin{equation}\label{eq:b-extremal}
	\dot{\fc{V}}^i = \delta^{im} \fc{Q}^k\sb{j}\, \fc{R}^{j}\sb{klm} \fc{V}^l,
\end{equation}
where the dot denotes a derivative with respect to $t$ and 
$\fc{Q}^i\sb{k}(t)$ is the solution of the initial value problem
\begin{equation}\label{eq:Qdef}
	\dot{\fc{Q}}^i\sb{k} = \fc{V}^i \fc{V}^j \delta_{jk}, \qquad \fc{Q}^i\sb{k}(a)=0.
\end{equation}
\end{proposition}

\begin{proof}
With a slight abuse of notation, we consider $\lambda$ to be a 
1-parameter family of curves with variational parameter $u$, such that 
$u=0$ corresponds to the original curve.  The variational vector field 
$X:=\dd{u}$ is assumed to be smooth with boundary conditions 
$X(0,u):=0$ and $X(a,u):=0$.  Parallel propagation of $E_p$ along 
$\lambda$ for each fixed $u$ gives a frame field $E(t,u)$, and a 
coframe field $\theta(t,u)$ dual to $E(t,u)$.  We also denote a 
$t$-derivative by a dot and write $V$ for the tangent of $\lambda$.  
The \b{length} functional \eqref{def:b-length} can then be written as
\begin{equation}\label{eq:b-functional}
	l(\lambda,E_p) = \int_0^a \abs{\theta(V)} \,\id t.
\end{equation}
Note that if $\lambdabar$ is the horizontal lift of $\lambda$ for each 
fixed $u$, then the lift of $\theta$ coincides with the canonical 
1-form $\canform$, so \eqref{eq:b-functional} agrees with the 
definition \eqref{def:b-length-OM} of \b{length} for horizontal curves 
in $OM$.

Differentiating \eqref{eq:b-functional} and evaluating at $u=0$, we get
\begin{equation}
\begin{split}
		\idd{u} l(\lambda,E_p) 
		&= \int_0^a \abs{\theta(V)}^{-1} \delta_{ij} \theta^{i}(V) 
								\dd{u}\bigl(\theta^{j}(V)\bigr) \,\id t \\
		&= \int_0^a \fc{V}^i \bigl( 
										  (\cov{X}\theta^{j})(V) + \theta^{j}(\cov{X}V)
									  \bigr) \,\id t
\end{split}
\end{equation}
where we have used that $\abs{\theta(V)}=1$ since $\lambda$ is 
parameterised by \b{length} at $u=0$, and the indices $i,j,k,\dots$ 
denote components in the frame $E$.  Using that $V=\fc{V}^iE_i$ and 
$[V,X]=0$,
\begin{equation}
\begin{split}
		\idd{u} l(\lambda,E_p) 
		&= \int_0^a \delta_{ij} \fc{V}^i \bigl( 
					\fc{V}^k (\cov{X}\theta^{j})(E_k) + \theta^{j}(\cov{V}X)
				\bigr) \,\id t \\
		&= \int_0^a \delta_{ij} \fc{V}^i \bigl( 
					- \fc{V}^k \theta^{j}(\cov{X}E_k) + \idd{t}(\fc{X}^j)
				\bigr) \,\id t,
\end{split}
\end{equation}
since $\cov{X}\bigl(\theta^{j}(E_k)\bigr)=0$ and 
$\cov{V}\theta^{j}=0$. Rewrite $\theta^{j}(\cov{X}E_k)$ as an 
integral from 0 to $t$ and perform a partial integration on the 
second term. Then 
\begin{equation}\label{eq:ddulambda3}
	\idd{u} l(\lambda,E_p) 
	= - \int_0^a \delta_{ij} \fc{V}^i \fc{V}^k 
							 \int_0^t \fc{R}^{j}\sb{klm} \fc{V}^l \fc{X}^m \,\id \hat{t}
							 \,\id t
		- \int_0^a \delta_{ij} \dot{\fc{V}}^i \fc{X}^j \,\id t.
\end{equation}
To proceed further we define $\fc{Q}^i\sb{k}(t)$ as the solution of the 
initial value problem \eqref{eq:Qdef}.  We can then partially 
integrate the first term in \eqref{eq:ddulambda3}, which results in
\begin{equation}
	\idd{u} l(\lambda,E_p) 
	= \int_0^a \bigl( 
								\fc{Q}^i\sb{j} \fc{R}^{j}\sb{ilm} \fc{V}^l 
								- \dot{\fc{V}}^i \delta_{im} 
							\bigr) \fc{X}^m \,\id t.
\end{equation}
By the basic principle of variational calculus, we arrive at 
condition \eqref{eq:b-extremal}.
\end{proof}

Based on Proposition~\ref{pr:b-extremal}, we can make some remarks on 
\b{length} extremals:
\begin{enumerate}
	\item Given a value for $\fc{V}(a)$, \eqref{eq:b-extremal} and 
				\eqref{eq:Qdef} determine unique solutions for $\fc{V}$ and 
				$\fc{Q}$ on some interval $[a-\epsilon,a]$.  So any point $q$ 
				has a neighbourhood $\U$ such that a \b{length} extremal from 
				$p$ to $q$ exists for all $p\in\U$.
	\item The two equations \eqref{eq:b-extremal} and \eqref{eq:Qdef} 
				may be viewed as an integro-differential equation for 
				$\fc{V}$.  Thus the situation is qualitatively different from 
				that of geodesics in $M$, which are solutions to a single 
				system of 4 ordinary differential equations.  Alternatively, 
				reformulating the problem in $OM$ as to find horizontal curves 
				with extremal \b{length}, \eqref{eq:b-extremal} and 
				\eqref{eq:Qdef} may be viewed as a single system of 10 
				ordinary differential equations.  There is a clear analogy to 
				the system of 10 geodesic equations in $OM$.  Hence it is 
				probably more natural and convenient to study \b{length} 
				extremals in the frame bundle context.
	\item \label{rem:subinterval}
				Since $\fc{Q}(a)=0$, \eqref{eq:b-extremal} requires 
				$\dot{\fc{V}}(a)=0$.  So the acceleration $\cov{V}V$ has a 
				zero at the end point $t=a$.  It follows that the restriction 
				of a \b{length} extremal to a subinterval is \emph{not} a 
				\b{length} extremal in general, since that requires that the 
				acceleration $\cov{V}V$ vanishes at the endpoint of the 
				subinterval.  This is not surprising, as varying a curve on a 
				subinterval $[0,b]\subset[0,a]$ affects the frame $E$ not only 
				on $[0,b]$ but also on $[b,a]$.
	\item The choice of the initial frame $E_p$ is crucial, as is
				evident from \eqref{eq:b-extremal}.
\end{enumerate}

If $\lambda$ is a geodesic, the $\fc{V}^i$ are constant so \eqref{eq:Qdef} 
can be integrated, which results in
\begin{equation}
	\fc{Q}^i\sb{k}(t) = \fc{V}^i \fc{V}^j \delta_{jk} (t-a).
\end{equation}
Inserting this into condition \eqref{eq:b-extremal} in 
Proposition~\ref{pr:b-extremal} we obtain the following corollary.
\begin{corollary}\label{co:b-extremalgeo}
Let $\lambda$ and $E$ be as in Proposition~\ref{pr:b-extremal}.  Then 
$\lambda$ is a geodesic only if
\begin{equation}\label{eq:b-extremalgeo}
	\delta_{ij} \fc{V}^i \fc{R}^{j}\sb{klm} \fc{V}^k \fc{V}^l = 0.
\end{equation}
\end{corollary}
\noindent
Note that \eqref{eq:b-extremalgeo} is algebraic, so if it is violated 
at one point then it is also violated on any interval containing that 
point. 

We now turn to the case where $\nabla$ is the Levi-Civit\`a connection 
of a Lorentzian metric $g$, and the frame $E_p$ is chosen to be 
pseudo-orthonormal with respect to $g$.  Then the parallel frame $E$ 
is also pseudo-orthonormal along any of the trial paths, so for all 
vector fields $X$ and $Y$,
\begin{equation}
	\fc{X}^i \fc{Y}^j \delta_{ij} = g(X,Y) + 2\,g(E_0,X)\,g(E_0,Y),
\end{equation}
where $E_0$ is the timelike vector of the frame $E$.  By 
Corollary~\ref{co:b-extremalgeo}, a \b{length} extremal is a geodesic 
only if
\begin{equation}
	\fc{V}^i \fc{R}_{iklm} \fc{V}^k \fc{V}^l 
	+ 2\, \fc{V}^0 \fc{R}_{0klm} \fc{V}^k \fc{V}^l = 0.
\end{equation}
By the symmetries of the curvature tensor $R$, the first term 
vanishes, and if $\lambda$ is causal, $\fc{V}^0\ne0$.  Thus a causal 
\b{length} extremal is a geodesic only if
\begin{equation}\label{b-extremalgeo2}
	\fc{R}_{0klm} \fc{V}^k \fc{V}^l = 0.
\end{equation}
It is apparent that \eqref{b-extremalgeo2} may be satisfied for some 
choice of $E_p$ and violated for some other choice. We wish to 
investigate if it is possible to choose $E_p$ such that \emph{all} 
sufficiently short causal \b{length} extremals starting at $p$ are 
geodesics. Since the causal vectors span the whole tangent space, 
\eqref{b-extremalgeo2} shows that this is possible if and only if 
\begin{equation}\label{b-extremalgeo3}
	0 = \fc{R}_{0klm} - \fc{R}_{0lkm} 
		= -\fc{R}_{ml0k} + \fc{R}_{mk0l} 
		= \fc{R}_{m0kl},
\end{equation}
because of the curvature identities.  Clearly, \eqref{b-extremalgeo3} 
holds only if $\fc{R}_{0l}=0$, i.e., the Ricci tensor $\fc{R}_{ij}$ 
must be degenerate.  This is of course an exceptional case not 
satisfied by a generic spacetime.

Finally, the results in this section is obviously in conflict with 
Lemma~3 of \cite{Stahl:variational}, which states that a non-geodesic 
causal curve in spacetime cannot be a \b{length} extremal.  This claim 
is incorrect.  As outlined in the proof, any non-geodesic smooth curve 
$\lambda$ may be restricted to a subinterval where the acceleration is 
bounded away from zero, and the restriction of $\lambda$ cannot be a 
\b{length} extremal.  However, as we have noted in 
remark~\ref{rem:subinterval} above, this does not imply that the whole 
curve cannot be a \b{length} extremal.  What is shown in Lemma~3 of 
\cite{Stahl:variational} is in fact that the acceleration cannot be 
bounded away from zero on a \b{length} extremal.  The reason is, as we 
have seen, that the acceleration must have a zero at the end point.

It follows that Theorem~2 of \cite{Stahl:variational} is incorrect as 
well.  If $M$ admits a covariantly constant timelike vector field 
$E_0$ with $g(E_0,E_0)=-1$, Lemma~3 and Theorem~2 may be 
reestablished, but that is a non-generic situation.

The remaining result of \cite{Stahl:variational}, Theorem~3, may be 
reestablished by other means, which we will do in 
section~\ref{sec:cluster}.

\section{Cluster curves}
\label{sec:cluster}

This section is devoted to the study of cluster curves, the main goal 
being to reestablish Theorem~3 of \cite{Stahl:variational}, which 
states that a partially imprisoned incomplete endless curve has an 
endless null geodesic cluster curve (see Theorem~\ref{th:imprisoned} 
below).  First we need a technical result, which will also be used in 
section~\ref{sec:nullcone}.

\begin{lemma}\label{la:cluster}
Let $\U\subset M$ and suppose that $p\in\U$ is a cluster point of a 
family of incomplete endless curves $\{\lambda_{i}\}$ in $\U$, with 
horizontal lifts $\{\lambdabar_{i}\}$ satisfying 
$l(\lambdabar_{i})\to0$ as $i\to\infty$.  If $\{\lambda_{i}\}$ has no 
subsequence that converges to a point in the topological closure 
$\cl\U$, then there is an inextendible null geodesic cluster curve of 
$\{\lambda_i\}$ through $p$ in $\cl\U$.
\end{lemma}

\begin{proof}
We may assume that $\lambdabar_i\colon[0,1)\to OM$.  Suppose that 
$\{\lambdabar_i\}$ has a cluster point $y\in OM$.  Then there is a 
sequence $\{t_j\}$ of real numbers such that\linebreak
$y_j:=\lambdabar_j(t_j)\to y$.  Let $\V$ be an arbitrarily small 
neighbourhood of $\pi(y)$ in $M$.  Then there is a small ball $B_r(y)$ 
around $y$ in $OM$ such that $\pi(B_r(y))\subset\V$.  But 
$l(\lambdabar_i)\to0$, so there is an $N\in\naturals$ such that 
$\lambdabar_j$ is contained in $B_r(y)$ for all $j\ge N$.  Then 
$\lambda_j$ is contained in $\V$ for all $j\ge N$, so $\{\lambda_j\}$ 
converges to $\pi(y)$ which contradicts the assumption on 
$\{\lambda_{i}\}$.

Since $p$ is a cluster point of $\{\lambda_i\}$, there is a sequence 
$\{t_j\}$ such that\linebreak $p_j:=\lambda_j(t_j)\to p$.  Put 
$\pbar_{\!j}:=\lambdabar_j(t_j)$.  By the argument in the previous 
paragraph, $\{\pbar_{\!j}\}$ has no cluster point in $OM$.  Let $\V$ be a 
convex normal neighbourhood of $p$ in $M$, let $\sigma\colon\V\to OM$ 
be a cross-section of $OM$ over $\V$, and let 
$\tilde\lambda_j(t):=\sigma\circ\lambda_j(t)$ whenever 
$\lambda_j(t)\in\V$.  The action of $\L$ on $OM$ is free and 
transitive, so there are unique matrices $\fc{L}_j(t)\in\L$ such that in 
$\pi^{-1}(\V)$,
\begin{equation}
	\lambdabar_j(t) = \tilde\lambda_j(t) \fc{L}_j(t).
\end{equation}
$\fc{L}_j(t)$ may be decomposed as 
$\fc{L}_j(t)=\OMEGAbar_j(t)\LAMBDA_j(t)\OMEGA_j(t)$, where 
$\OMEGA_j(t)$ and $\OMEGAbar_j(t)$ are spatial rotations and
\begin{equation}
	\LAMBDA_j(t) :=
	\begin{bmatrix}
		\cosh\xi_j(t) & \sinh\xi_j(t) & 0 & 0  \\
		\sinh\xi_j(t) & \cosh\xi_j(t) & 0 & 0  \\
		0 & 0 & 1 & 0  \\
		0 & 0 & 0 & 1  
	\end{bmatrix}
\end{equation}
is a Lorentz boost by a hyperbolic angle $\xi_j(t)\in\reals$.  Let 
$\xibar_j:=\xi_j(t_j)$.  If $\abs{\xibar_j}$ had an upper bound 
$\xi_0<\infty$, then $\{\pbar_{\!j}\}$ would be contained in a compact 
subset of $OM$, which is impossible since $\{\pbar_{\!j}\}$ has no 
cluster point.  We therefore assume that $\sup\{\xibar_j\}=\infty$, 
the case when $\inf\{\xibar_j\}=-\infty$ being similar.

Now $O(3)$ is compact, so there is a subsequence $\{\pbar_{\!k}\}$ of 
$\{\pbar_{\!j}\}$ such that\linebreak $\xi_k\to\infty$, 
$\OMEGA_k(t_k)\to\OMEGA_0$ and 
$\OMEGAbar_k(t_k)\to\OMEGAbar_0$ as $k\to\infty$.  Let
\begin{equation}
	\lambda_k'(t) := \lambdabar_k(t)\OMEGA_k(t_k)^{-1}
		= \tilde\lambda_k(t)\OMEGAbar_k(t)\LAMBDA_k(t)
			\OMEGA_k(t)\OMEGA_k(t_k)^{-1}
\end{equation}
and 
\begin{equation}
	\hat\lambda_k(t) := \lambdabar_k(t)\OMEGA_k(t_k)^{-1}\LAMBDA_k(t_k)^{-1}
		= \tilde\lambda_k(t)\OMEGAbar_k(t)\LAMBDA_k(t)
			\OMEGA_k(t)\OMEGA_k(t_k)^{-1}\LAMBDA_k(t_k)^{-1}.
\end{equation}
Then 
\begin{equation}
	\hat\lambda_k(t_k) = \tilde\lambda_k(t_k)\OMEGAbar_k(t_k)
	\to\hat{p}:=\sigma(p)\OMEGAbar_0
\end{equation}
as $k\to\infty$.  Since $\OMEGA_k(t_k)$ is a constant rotational 
matrix, leaving the Euclidian norm invariant, it does not affect the 
length of $\lambda_k'$.  From $l(\lambdabar_k)\to0$ it follows that 
$l(\lambda_k')\to0$ and so
\begin{equation}
	\int_{t_k}^1 \abs{\fc{X}_k^I} \,\id t \to 0, \qquad I=u,v,2,3,
\end{equation}
where $\fc{X}_k^I:=G(E_I,\dot\lambda_k')$, 
$E_u:=\tfrac1{\sqrt2}(E_0+E_1)$, $E_v:=\tfrac1{\sqrt2}(E_0-E_1)$ and 
$E_0$, $E_1$, $E_2$ and $E_3$ are the standard horizontal vector 
fields on $OM$ \cite{Kobayashi-Nomizu-I}.  Similarly, let 
$\fc{Y}_k^I=G(E_I,\dot{\hat\lambda}_k)$.  Then 
$\fc{Y}_k^u=e^{\xi_k}\fc{X}^u$, $\fc{Y}_k^v=e^{-\xi_k}\fc{X}^v$, 
$\fc{Y}_k^2=\fc{X}^2$ and $\fc{Y}_k^3=\fc{X}^3$, so
\begin{equation}\label{eq:Yv23bound}
	\int_{t_k}^1 \abs{\fc{Y}_k^I} \,\id t \to 0, \qquad I=v,2,3.
\end{equation}

Let $\mubar$ be the integral curve of $E_u$ through $\hat{p}$.  Then 
$\mu:=\pi\circ\mubar$ is a null geodesic in $\V$.  We may assume that 
$\mubar$ is extended as far as possible as the horizontal lift of an 
unbroken null geodesic in $\V$.  We show that $\mubar$ is a limit 
curve of $\{\hat\lambda_k\}$.  Let $q$ be a point on $\mubar$, let 
$\W$ be a neighbourhood of $q$ in $\pi^{-1}(\V)$, and let $\T$ be the 
tubular subset of $\pi^{-1}(\V)$ generated by all integral curves of 
$E_u$ intersecting $\W$.  Since $p\in\T$, \eqref{eq:Yv23bound} gives 
that there is an $N\in\naturals$ such that if $k>N$ then 
$\hat\lambda_k\cap\pi^{-1}(\V)$ is contained in $\T$, i.e., 
$\hat\lambda_k$ does not leave $\T$ except possibly at the ends 
$\dtop\bigl(\pi^{-1}(\V)\bigr)\cap\T$.  Now $\V$ does not contain any 
imprisoned incomplete curves since it is a convex normal 
neighbourhood, so $\hat\lambda_k$, having no endpoint in $\T\subset\V$, 
must leave $\T$.  Thus $\hat\lambda_k$ intersects $\W$ for each $k>N$, 
which means that $q$ is a limit point of $\{\hat\lambda_k\}$.

Obviously, $\mu$ is contained in $\cl\U$ since it is a limit curve of 
$\{\lambda_k\}$.  It remains to show that $\mu$ can be extended to an 
inextendible null geodesic cluster curve of $\{\lambdabar_i\}$ in the 
whole of\, $\cl\U$.  Extend $\mu$ as far as possible as an unbroken 
null geodesic in $\cl\U$, and let $q$ be a point on $\mu$.  Then the 
segment of $\mu$ from $p$ to $q$ is closed and finite, so it can be 
covered by a finite sequence of convex normal neighbourhoods 
$\{\V_n\}$ with $\V_1=\V$.  By the above argument, 
$\mu\cap\pi^{-1}(\V_1)$ is a limit curve of some subsequence 
$\{\lambdabar_k\}$ of $\{\lambdabar_i\}$.  Assume that 
$\mu\cap\pi^{-1}(\V_n)$ is a limit curve of a subsequence 
$\{\lambdabar_{k_n}\}$ for some $n$.  Then any point $p_n$ on 
$\mu\cap\pi^{-1}(\V_n)\cap\pi^{-1}(\V_{n+1})$ is a cluster point.  
Repeating the argument with $p_n$ in place of $p$ and $\V_{n+1}$ in 
place of\, $\V$ shows that $\mu\cap\pi^{-1}(\V_{n+1})$ is a limit 
curve of some subsequence $\{\lambdabar_{k_{n+1}}\}$ as well.  By 
induction, the whole curve $\mu$ is a cluster curve of 
$\{\lambdabar_i\}$.
\end{proof}

The proof of Lemma~\ref{la:cluster} uses a similar technique as the 
proof of the theorem in \cite{Schmidt:local-b-completeness}, except 
that we have weakened the assumption of total imprisonment to a family 
of curves with lengths going to 0, not converging to a point.  This 
allows us to use Lemma~\ref{la:cluster} in other contexts.  See also 
Proposition~8.3.2 in \cite{Hawking-Ellis}, but note that there are 
some minor errors in that version.

It is now a simple matter to apply Lemma~\ref{la:cluster} to 
imprisoned curves, which allows us to settle the issue from 
\cite{Stahl:variational} with the following theorem.

\begin{theorem}\label{th:imprisoned}
An incomplete endless curve partially imprisoned in a compact set 
admits an endless null geodesic cluster curve.
\end{theorem}

\begin{proof}
If $\lambda$ is an incomplete endless curve partially imprisoned in a 
compact set $\K$, then the intersection of $\lambda$ with the interior 
of $\K$ is a family of incomplete endless curves $\{\lambda_i\}$ with 
horizontal lifts whose lengths go to 0.  The problem is the endpoints 
of $\{\lambda_i\}$ on $\dtop\K$, and also the possibility that 
$\{\lambda_i\}$ contains subsequences converging to a point on 
$\dtop\K$ (see section~2).  But this can be dealt with by enlarging 
$\K$ around any such points.  Thus Lemma~\ref{la:cluster} gives us an 
inextendible null geodesic cluster curve $\gamma$ of $\{\lambda_i\}$ 
in $\K$.  Let $\mu$ be the endless extension of $\gamma$ as a null 
geodesic in $M$ and let $q$ be a point on $\mu$.  Then the segment of 
$\mu$ from $\K$ to $q$ is finite and so it can be included in a larger 
compact set $\K'$.  Applying Lemma~\ref{la:cluster} to $\K'$ we find 
that the part of $\mu$ in $\K'$ is a cluster curve of $\{\lambda_i\}$ 
as well, and since $q$ was arbitrary, the whole of $\mu$ is a cluster 
curve in $\K$.
\end{proof}

\section{\protect\b{neighbourhoods} and light cones}
\label{sec:nullcone}

In this section we will study how the light cone structure of $(M,g)$ 
is encoded in $(OM,G)$.  We also define a family of sets 
$\N_{p,\epsilon}(\U)$ that effectively describe neighbourhoods within 
a finite \b{distance} from a fixed point.  We start with a definition.

\begin{definition}\label{def:dtilde}
Given $\U\subset M$ and $p,q\in\U$, let 
\begin{equation}
	\dtilde_{\U}(p,q) := 
  \inf\{l(\mu);\; \mu\colon[0,1]\to\pi^{-1}(\U), 
									\pi\circ\mu(0)=p, 
									\pi\circ\mu(1)=q\}.
\end{equation}
\end{definition}
\noindent
$\dtilde_{\U}$ is not a metric on $\U$, since it is quite possible 
that $\dtilde_{\U}(p,q)=0$ with $p\ne q$.  Neither is it a semimetric 
in general, since the triangle inequality can be violated.  The 
case of interest is sets where $\dtilde$ is small, in the following 
sense:
\begin{definition}\label{def:Np}
Given $\U\subset M$, $p\in\U$ and $\epsilon>0$, let
\begin{equation}
	\N_{p,\epsilon}(\U) := \{q\in\U;\; \dtilde_{\U}(p,q) < \epsilon\}
\end{equation}
and
\begin{equation}
	\N_p(\U) := \{q\in\U;\; \dtilde_{\U}(p,q) = 0\}.
\end{equation}
\end{definition}
\noindent
It is clear from the definition of $\dtilde$ that
\begin{equation}
% 	\N_p(\U) = \underset{\raisebox{5pt}{$\scriptstyle{\epsilon>0}$}}\cap 
% 							\,\N_{p,\epsilon}(\U).
	\N_p(\U) = \bigcap_{\epsilon>0}\,\N_{p,\epsilon}(\U).
\end{equation}
Also, if $\epsilon_1<\epsilon_2$, 
$\N_{p,\epsilon_1}(\U)\subset\N_{p,\epsilon_2}(\U)$.  When 
$\U=M$, we will write $\N_p$ instead of $\N_p(M)$.

If $q\in\N_p(\U)$ there is a family of curves $\{\lambdabar_i\}$ from 
$\pi^{-1}(p)$ to $\pi^{-1}(q)$ in $OM$ such that $l(\lambdabar_i)\to0$ 
as $i\to\infty$.  We will refer to such a family of curves as 
\emph{defining} for $q\in\N_p(\U)$.  Because of 
Proposition~\ref{pr:horizontal} in Appendix~\ref{sec:horizontal}, we 
can assume that the $\lambdabar_i$ are horizontal.  In fact, we could 
have used horizontal curves from the outset with similar results:  if 
we replace $\dtilde$ with
\begin{equation}
	\dbar_{\U}(p,q) := 
  \inf\{l(\mu);\; \mu\colon[0,1]\to\pi^{-1}(\U), 
									\pi\circ\mu(0)=p, 
									\pi\circ\mu(1)=q, 
									\ver\dot\mu=0\}
\end{equation}
and $\N_{p,\epsilon}(\U)$ with
\begin{equation}
	\Nbar_{p,\epsilon}(\U) := \{q\in\U;\; \dbar_{\U}(p,q) < \epsilon\},
\end{equation}
then Proposition~\ref{pr:horizontal} implies that
\begin{equation}
	\N_{p,\epsilon}(\U) \subset \Nbar_{p,(e^\epsilon-1)}(\U).
\end{equation}
The sets $\Nbar_{p,\epsilon}(\U)$ are usually somewhat easier to 
work with since one can then restrict attention to horizontal curves.

\begin{proposition}\label{pr:lightcone}
The lightcone $N_p(\U)$, consisting of all points connected to $p$ 
by null geodesics in $\U$, is contained in $\N_p(\U)$.
\end{proposition}

\begin{proof}
Let $\gamma\colon[0,1]\to\U$ be a null geodesic from $p$ to $q$ with 
affine parameter $t$.  Pick a pseudo-orthonormal frame $E$ at $p$ such 
that $\dot\gamma=(a/\sqrt2)(E_0+E_1)$ at $p$, and parallel propagate 
$E$ along $\gamma$.  The length of $\gamma$ in the frame $E$ is
\begin{equation}
	l(\gamma,E) = \int_{0}^{1} a \,\id t = a.
\end{equation}
Now let $\fc{L}\in\L$ be a boost in the $E_0+E_1$ direction by hyperbolic 
angle $\xi$, i.e., 
\begin{equation}\label{eq:E0E1boost}
	\fc{L} := 
	\begin{bmatrix}
		\cosh\xi & \sinh\xi & 0 & 0  \\
		\sinh\xi & \cosh\xi & 0 & 0  \\
		0 & 0 & 1 & 0  \\
		0 & 0 & 0 & 1  
	\end{bmatrix}
\end{equation}
in the frame $E$.  Then the length of $\gamma$ in the frame $E\fc{L}$ is $a 
e^{-\xi}$, which tends to 0 as $\xi\to\infty$.
\end{proof}

Proposition~\ref{pr:lightcone} gives a characterisation of some of the 
points in $\N_p(\U)$.  Using Lemma~\ref{la:cluster}, we can also say 
the following about $\N_p(\U)$.

\begin{theorem}\label{th:generators}
$\N_p(\U)$ is generated by inextendible null geodesics in $\U$.
\end{theorem}

\begin{proof}
Let $q$ be a point in $\N_p(\U)\setminus\{p\}$ and let 
$\{\lambdabar_i\}$ be a defining family of curves for $q$.  Since we 
may remove any loops at $p$, the projections $\lambda_i$ to $M$ are 
incomplete and endless curves in $\U\setminus\{p\}$.  Also, $p\ne q$ 
so no subsequence of ${\lambda_i}$ converges to a point.  Thus 
Lemma~\ref{la:cluster} with $q$ in place of $p$ implies that there is 
an inextendible null geodesic cluster curve $\gamma$ of some 
subsequence $\{\lambda_k\}$ through $q$ in $\cl\U$.  Let 
$\gamma'$ be the inextendible segment of $\gamma$ through $q$ in $\U$.  
We show that $\gamma'$ is contained in $\N_p(\U)$.  Suppose that $r$ 
is a point on $\gamma'$ and let $\tilde\gamma$ be the segment of 
$\gamma'$ from $q$ to $r$.  We may assume that $\tilde\gamma$ is 
parameterised by an affine parameter $t$ ranging from $0$ to $a$.  As 
in the proof of Lemma~\ref{la:cluster}, there is a pseudo-orthonormal 
frame $E$ at $q$ in which the tangent of $\tilde\gamma$ is 
$E_u:=\tfrac1{\sqrt2}(E_0+E_1)$, and $\lambdabar_k$ intersects 
$\pi^{-1}(q)$ at the frame $F:=E\LAMBDA_k\OMEGA_k$.  Let $\gammabar_k$ 
be the horizontal lift of $\tilde\gamma$ to $F$.  Then the component 
vector $\fc{V}$ of $\dot{\gammabar}_k$ with respect to the standard 
horizontal vector fields on $OM$ is
\begin{equation}
	\fc{V} = \tfrac1{\sqrt2}\,\OMEGA_k^{-1}\LAMBDA_k^{-1}\!
	\begin{bmatrix} 1 \\ 1 \\ 0 \\ 0 \end{bmatrix}.
\end{equation}
Since $\OMEGA_k$ is a rotation leaving $E_u$ fixed and 
$\LAMBDA_k$ is a boost in the $E_u$ direction with hyperbolic 
angle $\xi_k$, the Euclidian norm of\, $\fc{V}$ is
\begin{equation}
	\abs{\fc{V}} = e^{-\xi_k},
\end{equation}
so the \b{length} of $\gammabar_k$ is $ae^{-\xi_k}$.  It follows 
that the 
concatenation of $\lambdabar_k$ and $\gammabar_k$ is a curve 
from $\pi^{-1}(p)$ to $\pi^{-1}(r)$ with length $l(\lambdabar_k) + 
ae^{-\xi_k}$, which tends to 0 as $k\to\infty$.
\end{proof}

The following theorem gives some idea of in which situations 
$\N_p(\U)$ can be expected to contain more than the light cone 
$N_p(\U)$.

\begin{theorem}\label{th:Npcompact}
If\, $\U$ is a compact subset of $M$ without totally imprisoned null 
geodesics, then $\N_p(\U)=N_p(\U)$ for any $p\in\U$.
\end{theorem}

\begin{proof}
Let $q\in\N_p(\U)\setminus\{p\}$ and let $\{\lambdabar_i\}$ be a 
defining family of curves for $q$.  Since $p\ne q$, no subsequence of 
$\{\pi\circ\lambdabar_i\}$ converges to a point.  By 
Lemma~\ref{la:cluster}, there is a null geodesic limit curve $\gamma$ 
of a subsequence $\{\pi\circ\lambdabar_k\}$ of 
$\{\pi\circ\lambdabar_i\}$ through $q$ in $\U$ which is inextendible 
in $\U\setminus\{q\}$.  We assume that $\gamma$ never reaches $p$ and 
show that this leads to a contradiction.

Since there are no totally imprisoned null geodesics in $\U$, $\gamma$ 
must have an endpoint $r$ on $\dtop\U$.  Then $r$ is a limit point of 
$\{\lambdabar_k\}$.  Let $\V$ be a convex normal neighbourhood of $r$, 
sufficiently small for $p$ and $q$ not to be in $\V$.  Each curve 
$\lambdabar_k$ must enter and leave $\V$ for large enough $k$.  By 
Lemma~\ref{la:cluster} there is an endless null geodesic cluster curve 
of $\{\lambdabar_k\}$ through $r$ in $\V$.  But every $\lambdabar_k$ 
is contained in $\U$, so the cluster curve cannot leave $\U$.  We have 
thus obtained an extension of $\gamma$ in $\U$, which contradicts that 
$\gamma$ is inextendible.
\end{proof}

Theorem~\ref{th:Npcompact} is not entirely satisfactory, since the 
situation we are most interested in is when the closure 
$\overline{\U}$ in $\clbM$ contains points on the \b{boundary} $\dbM$.  
That is not possible if $\U$ is open and contains no imprisoned null 
geodesics.  To get some feeling of what to expect, we give some 
examples in the following section.

\section{Examples}
\label{sec:examples}

\subsection{Minkowski spacetime}
\label{sec:Minkowski}

\begin{figure}
\hspace*{\fill}
\begin{tabular}{@{}c@{}}
	\includegraphics{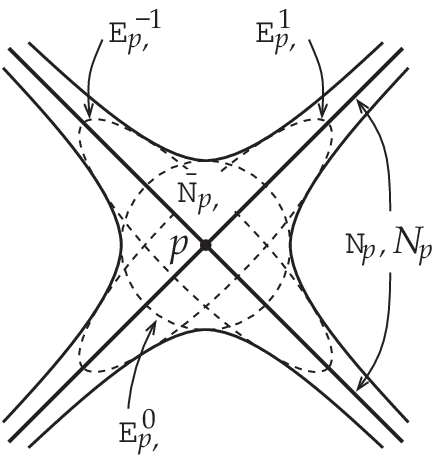}
\end{tabular}
\hfill\hfill
\begin{tabular}{@{}c@{}}
	\includegraphics{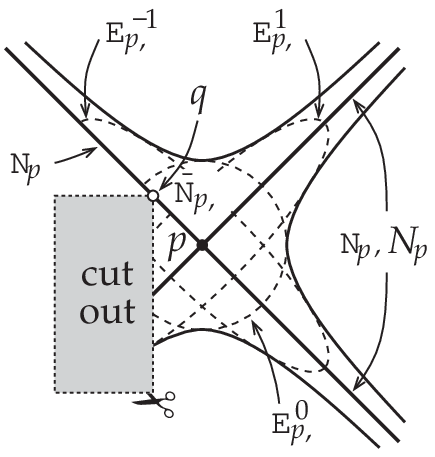}
	\hspace*{-4mm}
\end{tabular}
\hspace*{\fill}
\caption{
	The structure of $\Nbar_{p,\epsilon}$ in Minkowski spacetime 
	$\mathbb{M}$.  Three of the ellipsoidal sets 
	$\E_{p,\epsilon}^{\,\xi}$ are displayed, for $\xi=-1$,$0$ and $1$.  
	To the right a subset of $\mathbb{M}$ is cut out, showing how the 
	extension of a null geodesic which was originally passing through a 
	boundary point of the cut out set is recovered in $\N_p$.}
\protect\label{fig:Minkowski1}
\end{figure}

In Minkowski spacetime $\mathbb{M}$, the situation is of course very 
simple.  Let $\mathbb{M}=\reals^4$ with coordinates $(t,x,y,z)$ and 
line element
\begin{equation}
	\id s^2 = -\id t^2 + \id x^2 + \id y^2 + \id z^2. 
\end{equation}
Then the frame $E\fc{L}$ with $E=(-\dd{t},\dd{x},\dd{y},\dd{z})$ and 
constant $\fc{L}\in\L$ is parallel along any curve.  By symmetry we 
only need to consider a timelike 2-plane, $(t,x)$ say, with $\fc{L}$ a 
boost in that plane.  Suppose that a curve $\lambda$ is given by 
$\lambda(s)=(t(s),x(s))$, with frame $E\fc{L}$.  Introduce null 
coordinates $u:=\tfrac1{\sqrt{2}}(t+x)$ and 
$v:=\tfrac1{\sqrt{2}}(t-x)$, and assume that $\lambda$ is 
parameterised by \b{length} $s$.  Then there is a number $\xi$, the 
hyperbolic angle corresponding to $\fc{L}$, such that
\begin{equation}\label{eq:boost}
	E\fc{L} = (e^\xi\,\dd{u},e^{-\xi}\,\dd{v}),
\end{equation}
and the \b{length} functional is
\begin{equation}
	s = l(\lambda) = \int 
		\bigl( e^{-2\xi} \dot{u}^2 + e^{2\xi} \dot{v}^2 \bigr)^{1/2} 
		\,\id s.
\end{equation}
Since the integrand is functionally independent of $u$ and $v$, 
$\dot{u}$ and $\dot{v}$ must be constant on a curve with extremal 
\b{length}.  So for an extremal curve,
\begin{equation}
	s^2 = e^{-2\xi} u^2 + e^{2\xi} v^2.
\end{equation}
It follows that the set of points reachable on horizontal curves of 
length less than $\epsilon$ from the point $p$ with $(u,v)=(0,0)$ and 
frame given by $\xi$ is an ellipse of the form
\begin{equation}
	\E_{p,\epsilon}^{\,\xi} := 
		\{(u,v);\; e^{-2\xi} u^2 + e^{2\xi} v^2 < \epsilon^2 \}
\end{equation}
(see Figure~\ref{fig:Minkowski1}).  The structure in full Minkowski 
spacetime can then be found by applying spacelike rotations, giving 
ellipsoids in place of ellipses.  Note that
\begin{equation}
% 	\Nbar_{p,\epsilon} = 
% 		\underset{\raisebox{5pt}{$\scriptstyle\xi$}}\cap\, 
% 		\E_{p,\epsilon}^{\,\xi},
	\Nbar_{p,\epsilon} = \bigcap_\xi\, \E_{p,\epsilon}^{\,\xi},
\end{equation}
so we have a complete characterisation of $\Nbar_{p,\epsilon}$ 
(and hence a characterisation of $\N_{p,\epsilon}$ for small 
$\epsilon$). In particular, $\N_p=N_p$ for Minkowski spacetime.

To illustrate the importance of $\U$ being compact in 
Theorem~\ref{th:Npcompact}, we consider a modification of Minkowski 
spacetime by cutting out points.  If a point $q$ on one of the null 
geodesics from a point $p$ is cut out, the light cone $N_p$ will not 
contain the part of the null geodesic after the missing point.  But it 
will be contained in $\N_p$, provided that not too many points are 
missing around $q$ (see Figure~\ref{fig:Minkowski1}).

\subsection{Misner spacetime}
\label{sec:Misner}

Since the conditions of Theorem~\ref{th:Npcompact} exclude the case 
when $\U$ contains imprisoned null geodesics, it is interesting to 
study an example when this is the case.  We choose the Misner 
spacetime \cite{Misner:Taub-NUT,Hawking-Ellis,Ellis-Schmidt:singular}, 
as it is a simple example with imprisoned curves.

\begin{figure}
\begin{tabular}{@{}c@{}}
	\includegraphics{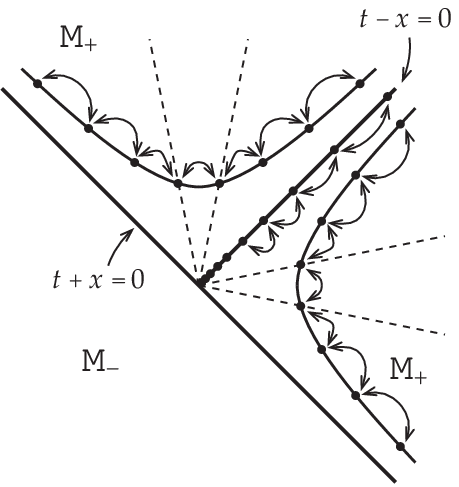}
	\vspace*{10pt}%\hspace*{2mm}
\end{tabular}
\begin{tabular}{@{}c@{}}
	\small Identify under $\G$\\
	\includegraphics{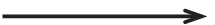}
\end{tabular}
\begin{tabular}{@{}c@{}}
	\hspace*{-6mm}
	\includegraphics{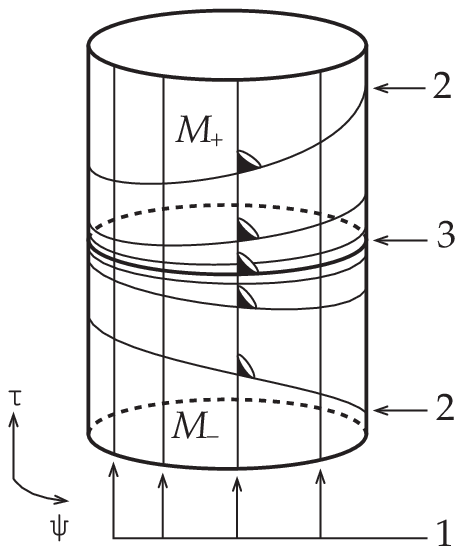}
\end{tabular}
\caption{
	Two-dimensional Misner spacetime.  Points in $\mathbb{M}_+$ are 
	identified under the discrete isometry group $\G$.  The arrows in 
	the left figure show the identification of points on orbits of 
	$\G$.  The figure on the right shows the three families of null 
	geodesics: family~1 obtained from null geodesics in $\mathbb{M}_+$ 
	with constant $t+x$, family~2 obtained from null geodesics with 
	constant nonvanishing $t-x$, and family~3 consisting of the single 
	null geodesic corresponding to $t-x=0$.}
\protect\label{fig:Misner1}
\end{figure}

Misner spacetime may be obtained from Minkowski spacetime 
$\mathbb{M}$ by identification under the isometry group generated by a 
fixed Lorentz boost $L_0$.  For simplicity, we restrict attention to 
two dimensions.  Let $L_0$ be given by
\begin{equation}
	L_0\colon (t,x) \mapsto 
		(t\cosh\xi_0 + x\sinh\xi_0, t\sinh\xi_0 + x\cosh\xi_0),
\end{equation}
and identify points on $\mathbb{M}_+:=\{(t,x);\; t+x>0\}$ under the 
discrete isometry group $\G$ generated by $L_0$.  We then obtain a 
spacetime with topology $\reals\times\sphere$.  If we introduce new 
coordinates (c.f.~\cite{Hawking-Ellis})
\begin{equation}
	\tau := \tfrac14(t^2-x^2)
	\qquad\text{and}\qquad
	\psi := \ln(t+x)^2 - \ln 4,
\end{equation}
with $\tau\in\reals$ and $\psi\in[0,2\pi]$, the Minkowski metric 
transforms to
\begin{equation}
	\id s^2 = 2\,\id\tau\,\id\psi + \tau\,\id\psi^2.
\end{equation}
The null geodesics of $M$ can be divided into three families (see 
Figure~\ref{fig:Misner1}):
\begin{enumerate}
	\item Null geodesics obtained from the null geodesics with constant 
				$t+x$ in $\mathbb{M}_+$.  Being given by constant $\psi$, they 
				are complete and pass through $\tau=0$.
	\item Null geodesics obtained from the null geodesics with constant
				$t-x$ in $\mathbb{M}_+$.  They are incomplete and endless, 
				spiralling around the spacetime indefinitely as $\tau\to0$.  
				Hence they are totally imprisoned in any neighbourhood of 
				$\tau=0$.
	\item The closed null geodesic at $\tau=0$, which is incomplete and 
				endless.
\end{enumerate}

\begin{figure}
% \hspace*{\fill}
\begin{tabular}{@{}c@{}}
	\includegraphics[scale=1.3]{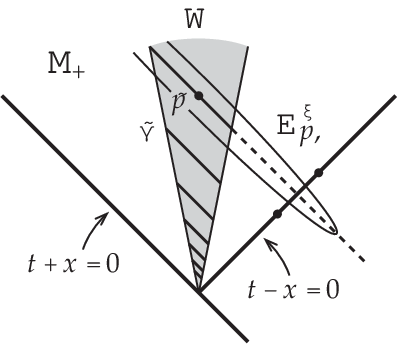}
\end{tabular}
\hfill
\begin{tabular}{@{}c@{}}
	\includegraphics[scale=1.3]{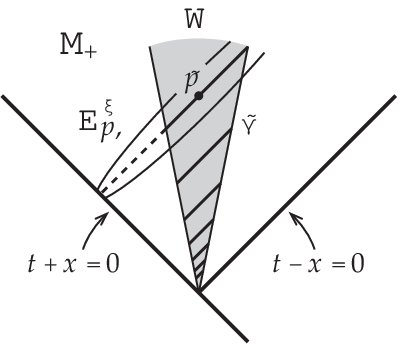}
\end{tabular}
% \hspace*{\fill}
\caption{
	A neighbourhood $\E_{p,\epsilon}^{\,\xi}$ in two-dimensional Misner 
	spacetime.  In the left figure, a part of a null geodesic of 
	family~1 is shown.  For sufficiently large negative $\xi$, no points 
	on the intersection of $\E_{p,\epsilon}^{\,\xi}$ with $t-x=0$ are 
	identified under $\G$.  On the right, it is shown how the past part 
	of a null geodesic of family~2 is contained in 
	$\E_{p,\epsilon}^{\,\xi}$ for large enough $\xi$.}
\protect\label{fig:Misner2}
\end{figure}

The structure of $\N_p$ for Misner spacetime may be deduced from our 
knowledge of the Minkowski case.  Let $M_+$ be the part of $M$ where 
$\tau>0$, and suppose that $p\in M_+$.  Also, let $L$ be a Lorentz 
boost with hyperbolic angle $\xi$.  We may identify $M_+$ with the 
wedge
\begin{equation}
	\W := \{(t,x)\in\mathbb{M};\; \abs{x/t} < \tanh(\xi_0/2),\, t>0 \}
\end{equation}
(see Figure~\ref{fig:Misner2}).  Let $\tilde{p}$ be the point in $\W$ 
corresponding to $p$, let $\gamma$ be a null geodesic through $p$, and 
let $\tilde{\gamma}$ be the corresponding null geodesic segments in 
$\W$ as in Figure~\ref{fig:Misner2}.  Clearly, the ellipsoidal 
neighbourhood $\E_{p,\epsilon}^{\,\xi}$ of $p$ in $\mathbb{M}$ 
corresponds to a neighbourhood of $\tilde{\gamma}$.  It follows that 
$\N_p(M_+)=N_p(M_+)$.  By a similar argument, the same holds for the 
part of $M$ with $\tau<0$.

We now include the set $\tau=0$.  We have two cases.  Suppose that 
$\gamma$ belongs to family~2, i.e., the extension of $\tilde{\gamma}$ 
passes through the line $t-x=0$ in $\mathbb{M}$.  As $\xi\to-\infty$, 
the intersection of $\E_{p,\epsilon}^{\,\xi}$ with $t-x=0$ tends to the 
intersection point of $\tilde{\gamma}$ with $t-x=0$, which of course 
corresponds to the intersection of $\gamma$ with $\tau=0$.

On the other hand, suppose that $\gamma$ belongs to family~1, i.e., 
the extension of the part of $\tilde{\gamma}$ through $p$ hits the 
line $t+x=0$ in $\mathbb{M}$.  Let $\gamma'$ be a null geodesic 
parallel to $\gamma$, with image $\tilde{\gamma}'$ in $\W$, such that 
$\tilde{\gamma}$ and $\tilde{\gamma}'$ are different null geodesics in 
$M$.  For large enough $\xi_0$, $\E_{p,\epsilon}^{\,\xi}$ does not 
intersect the extensions of the two segments of $\tilde{\gamma}'$ 
closest to $p$.  Since the isometry group is properly discontinuous, 
the same holds for the image of $\E_{p,\epsilon}^{\,\xi}$ and 
$\gamma'$ in $M$.  Hence no horizontal curve of sufficiently short 
\b{length} will reach $\tau=0$ in this direction, so the only points 
of $\N_p$ obtained in this way is the null geodesic $\gamma$ itself.

It remains to consider points $p$ lying on the closed null geodesic at 
$\tau=0$.  Suppose that there is a point $q\in\N_p$ which does not lie 
on $\tau=0$.  Then $p\in\N_q$, and we showed above that $\N_q=N_q$ if 
$\tau\ne0$ at $q$.  So there is a null geodesic from $q$ to $p$.  We 
conclude that $\N_p=N_p$ for Misner spacetime.

\subsection{Robertson-Walker spacetimes}
\label{sec:FLRW}

Of course, Misner spacetime might be uninteresting from a 
cosmological point of view, partly because it is flat, and partly 
because some cosmologists argue that there are no signs of topological 
pathologies in the real universe.  It is therefore important to obtain 
some results for more realistic cosmological models.  Some of the 
simplest are the Robertson-Walker models, with topology 
$M=I\times\Sigma$ where $I$ is a real interval, and line element
\begin{equation}
	\id s^2=-\id t^2+a(t)^2\id\sigma^2,
\end{equation}
such that $(\Sigma,\id\sigma^2)$ is a homogeneous space (see, e.g.,
\cite{Hawking-Ellis,Misner-Thorne-Wheeler}).  The scale function 
$a(t)$ is determined from the chosen matter model via Einstein's field 
equations.  For a Friedman big bang model, $a(t)\to0$ as $t\to0$, 
corresponding to a curvature singularity at $t=0$.

It can be shown that the \b{boundary} $\dbM$ is a single point 
\cite{Clarke:analysis-sing, Stahl:degeneracy, Bosshard:b-boundary, 
Johnson:b-boundary}.  Hence all null geodesics end at the same 
boundary point, and since the \b{length} of a null geodesic can be 
made arbitrarily small by an appropriate boost of the frame, the 
boundary point is not Hausdorff separated from any interior point of 
$M$.  Moreover, the boundary fibre in $OM$ is completely degenerate so 
the boundary of $OM$ is a single point as well.  This means that, 
along a curve ending at the singularity, choosing a different frame 
makes no difference at the boundary.  It would therefore seem like 
$\N_p$ should be the whole spacetime $M$.  But this is not necessarily 
the case, since the boundary point in $OM$ is singular with respect to 
the geometry in $OM$ as well, the curvature scalar tending to 
$-\infty$ at the boundary \cite{Stahl:LMgeometry}.

We will try to obtain an estimate of the neighbourhoods 
$\E_{p,\epsilon}^{\,\xi}$ for a particular Robertson-Walker model, 
valid for sufficiently small $\epsilon$.  However, we should mention 
from the outset that the estimates break down when 
$\E_{p,\epsilon}^{\,\xi}$ intersects the singularity.  This is 
unfortunate because the structure near the singularity is exactly what 
might cause identifications, giving a nontrivial $\N_p$.  The problem 
is the usual one when working with \b{length}: the length functional 
is not additive, in the sense that the length on a segment of the 
curve depends on the frame, which in turn is determined by parallel 
propagation along the \emph{whole} curve.  Also, the situation is not 
as simple as in Minkowski or Misner space, since the \b{extremal} 
curves are likely to develop `conjugate points'.

We will restrict attention to a two-dimensional model for simplicity.  
This is in fact not a restriction since $(\Sigma,\id\sigma^2)$ is 
homogeneous.  Let the metric be given by
\begin{equation}\label{eq:ds2FLRW}
	\id s^2 = -\id\tbar^2 + a(\tbar)^2\,\id x^2.
\end{equation}
We will fix the scale factor $a(\tbar)$ later.  If we replace the 
coordinate $\tbar$ with a conformal coordinate $t:=\int a^{-1}\id t$ 
and redefine the scale factor as a function $a(t)$ of $t$, the metric 
takes the form
\begin{equation}
	\id s^2 = a(t)^2 ( -\id t^2 + \id x^2 ).
\end{equation}
Any pseudo-orthonormal frame over $M$ may be expressed as $E\fc{L}(\xi)$, 
where $E$ is the global frame field given by 
$(a^{-1}\dd{t},a^{-1}\dd{x})$ and $\fc{L}(\xi)$ is a boost in the 
$t+x$ direction, i.e.,
\begin{equation}
	E\fc{L}(\xi) = (\cosh\xi\,E_0+\sinh\xi\,E_1,\sinh\xi\,E_0+\cosh\xi\,E_1).
\end{equation}

Let $\gamma$ be a horizontal curve given by $(t(s),x(s),\xi(s))$. 
The equation for parallel propagation of $\xi$ along $\gamma$ is
\begin{equation}\label{eq:xipp}
	\dot{\xi} + a'a^{-1} \dot{x} = 0,
\end{equation}
where $a'$ denotes the derivative of $a$ with respect to $t$.  
Expressing the tangent of $\gamma$ in the parallel frame $E\fc{L}(\xi)$ 
and inserting into the \b{length} formula \eqref{eq:b-functional} gives
\begin{equation}\label{eq:blengthgamma}
	l(\gamma) = 
		\int a \bigl(
			\dot{t}^2 \cosh 2\xi - 2\,\dot{t}\dot{x} \sinh 2\xi + \dot{x}^2 \cosh 2\xi
		\bigr)^{1/2} \,\id s.
\end{equation}
If we parameterise $\gamma$ by \b{length} $s$, we get
\begin{equation}\label{eq:bparam}
	\dot{t}^2 \cosh 2\xi - 2\,\dot{t}\dot{x} \sinh 2\xi + \dot{x}^2 \cosh 2\xi
	= a^{-2}.
\end{equation}
Now we assume that $\gamma$ is an extremal curve with respect to 
\b{length}.  Since the integrand of \eqref{eq:blengthgamma} is 
functionally independent of $x$, the functional derivative with 
respect to $\dot{x}$ gives a first integral
\begin{equation}\label{eq:firstint}
	\dot{x} \cosh 2\xi - \dot{t} \sinh 2\xi = \tfrac12 a^{-2} A,
\end{equation}
where $A$ is a constant determined by the initial values at $s=0$.  It 
is convenient to introduce an angular parameterisation of the initial 
values.  First, we define null coordinates on $M$ by
\begin{equation}\label{eq:uvdef}
	u := t + x \qquad\text{and}\qquad v := t - x.
\end{equation}
Then we may parameterise the initial conditions as 
\begin{equation}\label{eq:initconds}
	\dot{u}_0 := \sqrt{2}\, a_0^{-1} e^{\xi_0} \cos\theta
	\qquad\text{and}\qquad
	\dot{v}_0 := \sqrt{2}\, a_0^{-1} e^{-\xi_0} \sin\theta,
\end{equation}
where $\theta\in[0,2\pi)$ is a constant and $a_0:=a(t_0)$.  With this 
parameterisation $A$ is
\begin{equation}\label{eq:A}
	A = \sqrt{2}\, a_0 ( e^{-\xi_0} \cos\theta - e^{\xi_0} \sin\theta ).
\end{equation}

The three equations \eqref{eq:xipp}, \eqref{eq:bparam} and 
\eqref{eq:firstint} are sufficient for determining $\gamma$, given 
initial values for $t$, $x$, $\xi$ and $\theta$.  It is possible to 
solve \eqref{eq:firstint} for $\dot{x}$, and inserting the solution 
into \eqref{eq:bparam} we may solve for $\dot{t}^2$.  Put
\begin{equation}\label{eq:Wdef}
	W := a^2 \cosh 2\xi.
\end{equation}
Then \eqref{eq:xipp}, \eqref{eq:bparam} and \eqref{eq:firstint} are 
equivalent to the system
\begin{align}
	\label{eq:dt}
	\dot{t}^2 &= \tfrac14\,a^{-4}\,( 4W - A^2 ) \\
	\label{eq:dx}
	\dot{x} &= \frac{A}{2W} + \dot{t} \tanh 2\xi \\
	\label{eq:dxi}
	\dot{\xi} &= -a'a^{-1}\dot{x}.
\end{align}

For simplicity, we now restrict ourselves to the case when the scale 
factor is $a(\tbar)=\tbar^{1/2}$ (corresponding to a 
radiation-dominated universe), which will give us an idea about what 
to expect in general.  In the conformal coordinate~$t$, $a(t)=t/2$ and 
$a'(t)=1/2$.  We are now ready to state the result.

\begin{proposition}\label{pr:RW-nbhds}
Let $(M,g)$ be a two-dimensional Robertson-Walker spacetime with 
scale factor $a(\tbar)=\tbar^{1/2}$.  Let $\gamma$ be a curve of 
extremal \b{length}, parameterised by \b{length}~$s$ and starting at 
$(t_0,x_0,\xi_0)$.  Also, let $u=t+x$ and $v=t-x$.  Suppose that
$t<2t_0$ on $\gamma$. If $\xi_0>2$ then
\begin{displaymath}
	\abs{v-v_0} < 16\,a_0^{-1}e^{-\xi} s
\end{displaymath}
along $\gamma$.  On the other hand, if $\xi_0<-2$ then 
\begin{displaymath}
	\abs{u-u_0} < 16\,a_0^{-1}e^{\xi} s.
\end{displaymath}
\end{proposition}

\begin{proof}
We start by estimating $W$, given by \eqref{eq:Wdef}.  
Let $s_1$ be the largest number such that $W$ satisfies
\begin{equation}\label{eq:Wbounds}
	\frac12 < \frac{W}{W_0} < 2
\end{equation}
on $[0,s_1)$.  Here $W_0$ is the value of $W$ at $s=0$.  We show that 
either $t=0$ or $t=2t_0$ at $s=s_1$.

If we insert $a(t)=t/2$ in \eqref{eq:dt}, we get
\begin{equation}\label{eq:t2dt}
	\abs{t^2\dot{t}} = 2\sqrt{4W-A^2} \le 2\sqrt{4W}
\end{equation}
on $[0,s_1)$.  Using \eqref{eq:Wbounds} and integrating then gives
\begin{equation}\label{eq:t3t03}
	\abs{t^3-t_0^3} < 6\sqrt{8W_0}\,s.
\end{equation}

Next, from (\ref{eq:dt}--\ref{eq:dxi}) we have
\begin{equation}
	\idd{s}\sqrt{W^2-a^4} = - a'a^{-1}\!A = -t^{-1}\! A.
\end{equation}
Using \eqref{eq:t3t03} and integrating gives
\begin{equation}\label{eq:W2a4bound}
	\ABs{\sqrt{W^2-a^4}-\sqrt{\smash[b]{W_0^2-a_0^4}}}
	< \frac{\abs{A}}{4\sqrt{8W_0}}
		\Bigl( t_0^2 - \bigl(t_0^3 - 6\sqrt{8W_0}\,s\bigr)^{2/3} \Bigr)
	< \frac{\abs{A}}{4\sqrt{8W_0}}\, t_0^2.
	\raisetag{-3pt}
\end{equation}
Combining \eqref{eq:Wdef} and \eqref{eq:A}, we find that
\begin{equation}\label{eq:W0Abound}
	4W_0-A^2\ge0,
\end{equation}
so the right hand side of \eqref{eq:W2a4bound} is less than 
$t_0^2/5$.  Solving \eqref{eq:W2a4bound} for $W^2$ and dividing 
by $W_0^2$ we get
\begin{equation}
	\frac{W^2}{W_0^2} 
	< \frac{a^4}{W_0^2} 
		+ \biggl( \frac{\sqrt{\smash[b]{W_0^2-a_0^4}}}{W_0} 
					 + \frac{t_0^2}{5W_0}
			\biggr)^2.
\end{equation}
Since $t<2t_0$ and $\abs{\xi_0}>2$,
\begin{equation}\label{eq:Wupperbound}
	\frac{W}{W_0} < 1.1 < 2.
\end{equation}
Going back to \eqref{eq:W2a4bound} and estimating from below results 
in
\begin{equation}
	\frac{W^2}{W_0^2} 
	> \frac{a^4}{W_0^2} 
		+ \biggl( \frac{\sqrt{\smash[b]{W_0^2-a_0^4}}}{W_0} 
					 - \frac{t_0^2}{5W_0}
			\biggr)^2.
\end{equation}
The first term is positive, and expanding the square and applying the
conditions on $t$ and $\xi_0$ gives
\begin{equation}\label{eq:Wlowerbound}
	\frac{W}{W_0} > 0.98 > \frac12.
\end{equation}

From \eqref{eq:Wupperbound} and \eqref{eq:Wlowerbound} it follows that 
\eqref{eq:Wbounds} cannot be violated even at $s=s_1$.  So unless 
$t(s_1)=0$, the only remaining possibility is that $t(s_1)=2t_0$.

The next step is to estimate $\xi$ in terms of $\xi_0$.  Using the
definition \eqref{eq:Wdef} of $W$ and the lower bound of 
\eqref{eq:Wbounds} gives
\begin{equation}
	e^{2\abs\xi} 
	> \cosh 2\xi 
	> \frac{t_0^2}{2t^2} \cosh 2\xi_0
	> \frac{1}{16}\, e^{2\abs{\xi_0}},
\end{equation}
hence
\begin{equation}\label{eq:xibound}
	\abs{\xi} > \abs{\xi_0} - \ln 4.
\end{equation}

We can now provide bounds for $\dot{u}=\dot{t}+\dot{x}$ and 
$\dot{v}=\dot{t}-\dot{x}$.  Suppose that $\xi_0>2$.  
From \eqref{eq:dt} and \eqref{eq:dx},
\begin{equation}
	\abs{\dot{v}} 
	= \ABS{(1+\tanh 2\xi)\,\dot{t} + \frac{A}{2W}}
	< W^{-1/2} (e^{2\xi}+\sqrt2),
\end{equation}
where the inequality follows from \eqref{eq:W0Abound} and 
\eqref{eq:Wbounds}.  But
\begin{equation}
	W^{-1/2} (e^{2\xi}+\sqrt2)
	< 2\sqrt2\,a_0^{-1} e^{-\xi}\frac{\cosh\xi}{\sqrt{\cosh 2\xi}}
	< 2\sqrt2\,a_0^{-1} e^{-\xi},
\end{equation}
so using \eqref{eq:xibound} and integrating gives the desired bound on 
$\abs{v-v_0}$.  The argument for the case when $\xi_0<-2$ is similar.
\end{proof}

The problem when trying to use Proposition~\ref{pr:RW-nbhds} to 
estimate the extent of $\E_{p,\epsilon}^{\,\xi}$ is that while we have 
valid estimates for curves `near' $p$, it is likely that curves that 
approach $t=0$ will no longer have minimal \b{length}.  In a sense, 
there will be `conjugate points' with respect to the \b{length}.  Is 
it possible to find arbitrarily short curves between two distinct null 
geodesics?  It seems unlikely since boosting the frame in order to get 
close to the singularity will probably make it impossible to move a 
finite distance in the $x$-direction without spending too much 
\b{length}.  We therefore make the following conjecture.

\begin{conjecture}
In a Robertson-Walker spacetime, with a `physically reasonable' 
equation of state, $\N_p=N_p$.
\end{conjecture}

\section{Imprisonment and fibre degeneracy}
\label{sec:imp-fibre}

Let $\gamma:(0,1]\to OM$ be a horizontal curve with $\gamma(t)\to 
E\in\d OM$ as $t\to0$.  In \cite{Stahl:degeneracy} it was shown that 
if there are sequences $t_i\to 0$ and $\rho_i\to 0$ of real numbers 
such that the following conditions hold, the boundary fibre containing 
$E$ is totally degenerate:
\begin{enumerate}
	\item the closure of each ball $\U_i := B_{\rho_i}(\gamma(t_i))$ in 
				$\OMbar$ is compact and contained in $OM$.
	\item \label{cond:invertibility}
				$\fc{R}$, the frame components of Riemann tensor viewed as a 
				map from the space of bivectors to the Lie algebra, is invertible on 
				each $\U_i$.
	\item \label{cond:divergence}
				$\norm{\fc{R}(\gamma(t_i))^{-1}}^3 \sup_{\U_i}\norm{\fc{R}}^2$, 
				$\norm{\fc{R}(\gamma(t_i))^{-1}}^2 \sup_{\U_i}\norm{\fc{\cov R}}$ and 
				$\norm{\fc{R}(\gamma(t_i))^{-1}}/\!\rho_i$ all tend to 
				$0$ as $t_i \to 0$.  Here $\norm{\cdot}$ is the mapping norm 
				with respect to the frame in $OM$ and a fixed basis in the Lie 
				algebra, respectively.
\end{enumerate}
An explanation of condition~\ref{cond:invertibility} is given in 
Appendix~\ref{sec:invert}.  We will now investigate if 
condition~\ref{cond:divergence} is applicable to boundary points 
arising from imprisoned curves.

Suppose that an incomplete endless curve $\gamma$ is (partially or 
totally) imprisoned in a compact set $\K$.  If the spacetime is 
sufficiently general, in particular, if it is of Petrov type~I, some 
component of the Riemann tensor diverges in a parallel frame along 
$\gamma$ (see~\cite{Hawking-Ellis}, Proposition~8.5.2).  For 
condition~\ref{cond:divergence} to hold, it is necessary that 
$\norm{\fc{R}^{-1}}\to0$, which is true if and only if 
$\norm{\fc{R}(\fc{B})}\to\infty$ for all bivectors $\fc{B}$ with 
$\norm{\fc{B}}=1$~\cite{Stahl:degeneracy}.  So several components of 
$\fc{R}$ have to diverge in a specific manner.

Let $p\in\K$ be a cluster point of $\gamma$, let $\U$ be a convex 
normal neighbourhood around $p$ and let $\sigma$ be a section over 
$\cl\U$.  Then $\norm{\fc{R}}$ is bounded on $\sigma(\U)$ since 
$\sigma(\U)$ is contained in a compact set.  It is clear from the 
proof of Lemma~\ref{la:cluster} that a diverging $\fc{R}$ can only be 
caused by a diverging Lorentz transformation along~$\gamma$.

Since $\norm{\fc{R}^{-1}}$ is unaffected by spatial rotations, we only 
need to study the effect of a boost.  Fix a frame $E$ at a point 
$q\in\U$ and put $E_u:=(1/\sqrt2)(E_0+E_1)$ and 
$E_v:=(1/\sqrt2)(E_0-E_1)$.  Let $\fc{L}$ be a boost by an hyperbolic 
angle $\xi$ in the $E_u$ direction, as given by 
equation~\eqref{eq:E0E1boost}, and let $\fc{B}=E_u\wedge E_2$.  Then 
if $\fc{R}$ is the Riemann tensor expressed in the boosted frame 
$E\fc{L}$ and $R^i\sb{jkl}$ are the components of the Riemann tensor 
in the fixed frame $E$,
\begin{equation}
	\norm{\fc{R}(\fc{B})}^2 = (R^u\sb{2u2})^2 + (R^u\sb{3u2})^2 + o(e^{-2\xi}),
\end{equation}
where $o(e^{-2\xi})$ denotes terms less than a constant times 
$e^{-2\xi}$ for $\xi$ sufficiently large.  So there is a bivector 
$\fc{B}$ such that $\fc{R}(\fc{B})$ is bounded away from 0, which 
implies that $\norm{\fc{R}^{-1}}\not\to0$.  We conclude that 
condition~\ref{cond:divergence} does not hold for points in $\dOM$ 
arising from imprisoned curves.  Note that even though the techniques 
in \cite{Stahl:degeneracy} do not apply, the boundary fibre might 
still be partially or totally degenerate.

\section{Discussion}
\label{sec:discussion}

It seems likely that the compactness and non-imprisonedness conditions 
in Theorem~\ref{th:Npcompact} may be removed, at least in some cases.  
A first step would be to extend Proposition~\ref{pr:RW-nbhds} to cover 
more general Robertson-Walker spacetimes, and perhaps to other 
cosmological models.  That would give a better handle on \b{boundary} 
issues in more realistic cosmologies.  Also, in Schwarzschild 
spacetime it is still unknown if the \b{boundary} is a set of 
dimension~0 or 1.  Hopefully, the techniques used in 
section~\ref{sec:FLRW} can be generalised to cover a two-dimensional 
version of the Schwarzschild spacetime as well, since in two 
dimensions the inner part (i.e., inside the event horizon) can be 
written in the Robertson-Walker form \eqref{eq:ds2FLRW} for a 
particular choice of scale factor $a(\tbar)$.

It would also be interesting to obtain a fibre degeneracy theorem,
similar to the one in \cite{Stahl:degeneracy}, that is applicable to 
the imprisoned curve setting. It seems probable that only partial 
degeneracy can be expected in this case. However, very different 
techniques will be needed than those used in \cite{Stahl:degeneracy}.

\section*{Acknowledgements}

I am indebted to my thesis advisor Clarissa-Marie Claudel, in 
particular for some of the ideas explored in 
section~\ref{sec:nullcone} and section~\ref{sec:examples}, and to 
V.~Perlick of course, whose comments on variations of \b{length} I 
have paraphrased in section~\ref{sec:imp-variation}.

\appendix
\section{Horizontal curves}
\label{sec:horizontal}

When working in the pseudo-orthonormal frame bundle $OM$ it is often 
convenient to restrict attention to horizontal curves.  A statement of 
the following form can be found in the literature 
(cf.~\cite{Dodson:edge-geometry}, p.~442 and 
\cite{Clarke:analysis-sing}, pp.~36--38).

\begin{claim}
	Let $\tilde\lambda\colon[0,a)\to OM$ be a finite curve and let 
	$\lambdabar$ be the horizontal lift of $\pi\circ\tilde\lambda$ with 
	$\lambdabar(0)=\tilde\lambda(0)$.  Then 
\begin{equation}\label{eq:lambdaclaim}
		l(\lambdabar)\le l(\tilde\lambda).
\end{equation}
\end{claim}

However, this statement is generally false, as we will now see.  Given 
$\tilde\lambda$ and $\lambdabar$ as above, there is a curve $\fc{L}$ in 
$\L$ such that $\tilde\lambda(t)=\lambdabar(t)\fc{L}(t)$ for all 
$t\in[0,a)$, with $\fc{L}(0)=\delta$, the identity in $\L$.  Then
\begin{equation}
	\dot{\tilde\lambda}(t) 
% 	= \idd{t}\bigl( R_{\fc{L}(t)}(\lambdabar(t)) \bigr)
	= R_{\fc{L}(t)*}(\dot{\lambdabar}(t)) +
		\idd{s}\Bigr|_{s=t}\bigl( R_{\fc{L}(s)}(\lambdabar(t)) \bigr)
	= R_{\fc{L}(t)*}(\dot{\lambdabar}(t)) + \varphi(\fc{L}^{-1}\dot{\fc{L}}),
\end{equation}
where $\varphi$ is the canonical isomorphism from the Lie algebra 
$\l$ to the vertical subspace $V(OM)$ of\, $T(OM)$ at 
$\tilde\lambda(t)$ \cite{Kobayashi-Nomizu-I}. Now
\begin{equation}\label{eq:thetalambda}
	\canform(\dot{\tilde\lambda}) = \canform(R_{\fc{L}*}\dot{\lambdabar}) = 
	\fc{L}^{-1}\canform(\dot{\lambdabar})
\end{equation}
because of the transformation properties of the canonical 1-form 
$\canform$ under the right action of $\L$ \cite{Kobayashi-Nomizu-I}. 
Also
\begin{equation}\label{eq:omegalambda}
	\connform(\dot{\tilde\lambda}) = \varphi^{-1}(\ver \dot{\tilde\lambda}) 
	= \fc{L}^{-1}\dot{\fc{L}},
\end{equation}
where $\ver \dot{\tilde\lambda}$ is the vertical component of 
$\dot{\tilde\lambda}$ \cite{Kobayashi-Nomizu-I}.  We conclude that
\begin{equation}\label{eq:ltildelambda}
	l(\tilde\lambda) 
	= \int_0^a \Bigl( 
								\norm{\fc{L}^{-1}\canform(\dot{\lambdabar})}^2 
								+ \norm{\fc{L}^{-1}\dot{\fc{L}}}^2
						 \Bigr)^{1/2} \,\id t.
\end{equation}
It seems that the mistakes in~\cite{Clarke:analysis-sing} and 
\cite{Dodson:edge-geometry} stem from neglecting the $\fc{L}^{-1}$ factor, 
which originates from the \b{norm} being evaluated at different 
points in the fibre over $\pi\circ\tilde\lambda(t)$.

As an example, consider a null geodesic $\gamma$ with horizontal lift 
$\gammabar$, affinely parameterised by $t\in[0,a)$.  Let 
$\tilde\gamma$ be the curve given by 
$\tilde\gamma(t)=\gammabar(t)\fc{L}(t)$ where $\fc{L}(t)$ is a Lorentz boost in 
the direction of $\dot\gamma$ by an hyperbolic angle $\xi(t)$ with 
$\xi(0)=0$.  Then 
\begin{equation}
	l(\tilde\gamma)
	= \int_0^a \Bigl(	2\dot\xi(t)^2 + e^{-2\xi(t)} \Bigr)^{1/2} \,\id t,
\end{equation}
which certainly can be made smaller than $l(\gammabar)=a$ by an 
appropriate choice of $\xi(t)$.

However, note that in the frame bundle of a Riemannian manifold,
\begin{equation}
	\norm{\fc{L}^{-1}\canform(\dot{\lambdabar})}=\abs{\canform(\dot{\lambdabar})}
\end{equation}
since in that case $\fc{L}\in O(4)$, the orthogonal transformation 
group.  It follows that \eqref{eq:ltildelambda} reduces to
\begin{equation}
	l(\tilde\lambda) 
	= \int_0^a \Bigl( 
								\abs{\canform(\dot{\lambdabar})}^2 
								+ \norm{\fc{L}^{-1}\dot{\fc{L}}}^2
						 \Bigr)^{1/2} \,\id t
	\le \int_0^a \abs{\canform(\dot{\lambdabar})} \,\id t
	= l(\lambdabar).
\end{equation}
This result was used by Schmidt \cite{Schmidt:b-boundary} in proving 
that the \b{completion} is equivalent to the Cauchy completion in the 
Riemannian case.

In the Lorentzian case, it is still possible to find a connection 
between the lengths of horizontal curves and more general curves, 
being almost as strong as the relation \eqref{eq:lambdaclaim} for 
short curves.

\begin{proposition}\label{pr:horizontal}
Let $\tilde\lambda:[0,a)\to OM$ be a curve with finite \b{length}, and 
let $\lambdabar$ be the horizontal lift of $\pi\circ\tilde\lambda$ 
with $\lambdabar(0)=\tilde\lambda(0)$.  Then
\begin{equation}
	l(\lambdabar) \le e^{l(\tilde\lambda)}-1.
\end{equation}
\end{proposition}

\begin{proof}
We may assume that $\tilde\lambda$ is parameterised by \b{length} and 
that $\tilde\lambda(t)=\lambdabar(t)\fc{L}(t)$ for some curve $\fc{L}$ in $\L$, 
with $\fc{L}(0)=\delta$. Then by \eqref{eq:ltildelambda},
\begin{equation}\label{eq:tildelambdabound}
	\abs{\dot{\tilde\lambda}}^2 
	= \abs{\canform(\dot{\tilde\lambda})}^2 + \norm{\fc{L}^{-1}\dot{\fc{L}}}^2
	= 1,
\end{equation}
so $\abs{\canform(\dot{\tilde\lambda})}\le1$.  Since $\fc{L}(t)$ is a curve 
in $\L$, there is a curve $\mu$ in the Lie algebra $\l$ such that 
$\fc{L}(t)=\exp\mu(t)$, where $\exp$ is the exponential map $\l\to\L$ and 
$\mu(0)=0$.  Then by \eqref{eq:tildelambdabound}, 
$\abs{\dot\mu}=\norm{\fc{L}^{-1}\dot{\fc{L}}}\le1$.  It follows that
\begin{equation}
	\idd{t}\abs{\mu} \le \abs{\dot\mu} \le 1,
\end{equation}
which on integration gives $\abs\mu\le t$. Thus 
\begin{equation}
	\norm{\fc{L}} \le \abs{\exp\mu} \le e^{\abs\mu} \le e^t,
\end{equation}
so using \eqref{eq:thetalambda} gives
\begin{equation}
	l(\lambdabar) = \int_0^a \abs{\fc{L}\canform(\dot{\tilde\lambda})} \,\id t
	\le \int_0^a e^t \,\id t = e^a-1.
\end{equation}
\end{proof}

Proposition~\ref{pr:horizontal} then reestablishes the result in 
section~3.13 of \cite{Dodson:edge-geometry}, p.~441, and the crucial 
steps in the proof of Proposition~3.2.1 of 
\cite{Clarke:analysis-sing}, p.~38.

\section{Invertibility of the Riemann tensor}
\label{sec:invert}

This section serves to clarify the invertibility condition on the 
Riemann tensor in a given frame, viewed as a linear map from the space 
of bivectors to the Lie algebra of the Lorentz group.  Clearly, if 
$\fc{R}$ is invertible in one frame at a point $p\in M$, it is 
invertible in any other frame at $p$.  We restrict attention to the 
vacuum case when $\fc{R}=\fc{C}$, the Weyl tensor, for simplicity.  We 
will investigate the relation between invertibility of $\fc{C}$ and 
Petrov types, so we use a spinor formalism (see, 
e.g.,~\cite{Penrose-Rindler-I} and \cite{Stewart:advanced-gr}).  This 
requires a change of signature of the metric, which has no influence 
on the invertibility.  Also, we may study 
$\fc{C}_{abcd}=\eta_{ae}\fc{C}^{e}\sb{bcd}$ instead of 
$\fc{C}^{a}\sb{bcd}$.  In spinor form, we have
\begin{equation}
	\fc{C}_{abcd} = \fc{C}_{ABCDA'B'C'D'} 
					 = \Psi_{\!ABCD}\,\epsilon_{\!A'B'}\,\epsilon_{C'D'}
					 + \Psibar_{\!A'B'C'D'}\,\epsilon_{\!AB}\,\epsilon_{CD},
\end{equation}
where $\Psi_{\!ABCD}$ is the symmetric Weyl spinor. Any bivector 
$\fc{B}^{cd}$ may be decomposed as
\begin{equation}
	\fc{B}^{cd} = \fc{B}^{CDC'D'} 
				 = \phi^{CD}\epsilon^{C'D'} + \phibar^{\,C'D'}\epsilon^{CD}
\end{equation}
where $\phi^{CD}$ is a symmetric spinor.  Let $\mathcal{S}^2$ be the 
space of symmetric contravariant valence 2 spinors, and let 
$\mathcal{S}^{2*}$ be the dual space of $\mathcal{S}^2$.  It is 
easily found that $\fc{C}_{abcd}\fc{B}^{cd}=0$ for some bivector 
$\fc{B}^{cd}$ if and only if $\Psi_{\!ABCD}\,\phi^{CD}=0$ for some 
\hbox{$\phi^{CD}\in\mathcal{S}^2$}.  So $\fc{C}$ is invertible if and only 
if $\Psi\colon\mathcal{S}^2\to\mathcal{S}^{2*}$ is invertible.

Given a spin basis $\omicron^A,\iota^A$, we define a basis for 
$\mathcal{S}^2$ by
\begin{equation}
	E_{1}^{AB} = \omicron^A\omicron^B, \qquad
	E_{2}^{AB} = \omicron^{(A}\iota^{B)} \qquad
	\text{and}\qquad
	E_{3}^{AB} = \iota^A\iota^B.
\end{equation}
Then the corresponding dual basis for $\mathcal{S}^{2*}$ is
\begin{equation}
	E^{1}_{AB} = \iota_{\!A}\iota_B, \qquad
	E^{2}_{AB} = -\omicron_{(A}\iota_{B)} \qquad
	\text{and}\qquad
	E^{3}_{AB} = \omicron_{\!A}\omicron_B.
\end{equation}
In this basis, $\Psi:\mathcal{S}^2\to\mathcal{S}^2$ may be written as 
\begin{equation}
	\Psi =
	\begin{bmatrix}
		\Psi_0 & \Psi_1 & \Psi_2  \\
		\Psi_1 & \Psi_2 & \Psi_3  \\
		\Psi_2 & \Psi_3 & \Psi_4  
	\end{bmatrix}.
\end{equation}
Thus the determinant $\det{\Psi}$ is one of the two independent 
curvature scalars that can be constructed from the Weyl tensor.  Since 
the invertibility of $\Psi$ is independent of the choice of spin 
basis, we can choose $\omicron^A$ as one of the principal null 
directions of $\Psi_{\!ABCD}$.  Then $\Psi_0=0$, and the determinant of 
$\Psi$ becomes
\begin{equation}\label{eq:detPsi}
	\det{\Psi} 
	= -\Psi_2^3 + 2\,\Psi_1\Psi_2\Psi_3 - \Psi_1^2\Psi_4.
\end{equation}

Now if the Weyl tensor is of type III, N or O, three principal spinors 
of $\Psi$ coincide.  If we choose this repeated spinor as 
$\omicron^A$, $\Psi_1=\Psi_2=0$, so $\Psi$ is singular.  On the other 
hand, if only two principal spinors coincide, i.e., the Weyl tensor is 
of type II or D, $\Psi_1=0$ and $\Psi_2\neq0$ so $\Psi$ is invertible.

It remains to study the most general case with no repeated principal 
spinors (Petrov type I).  We may choose $\omicron^A$ and $\iota^A$ as 
two of the four principal spinors.  Then $\Psi_0=\Psi_4=0$, and we may 
write $\Psi_{\!ABCD}=\omicron_{(A}\iota_B\alpha_C\beta_{D)}$ for two 
linearly independent spinors $\alpha_{\!A}$ and $\beta_{\!A}$.  Let
\begin{equation}
	\alpha_{\!A} = \alpha_0 \omicron_{\!A} + \alpha_1 \iota_{\!A}
	\qquad\text{and}\qquad
	\beta_{\!A} = \beta_0 \omicron_{\!A} + \beta_1 \iota_{\!A}.
\end{equation}
Then 
\begin{equation}
	\Psi_1 = - \tfrac14 \alpha_1\beta_1,\quad
	\Psi_2 = \tfrac16 (\alpha_0\beta_1+\alpha_1\beta_0)
	\quad\text{and}\quad
	\Psi_3 = - \tfrac14 \alpha_0\beta_0.
\end{equation}
Now \eqref{eq:detPsi} is
\begin{equation}
	\det{\Psi} = -\Psi_2 ( \Psi_2^2 - 2\,\Psi_1\Psi_3 ),
\end{equation}
and the second factor is
\begin{equation}
	-\tfrac1{288} (\alpha_0\beta_1 + \alpha_1\beta_0)^2
	+ \tfrac1{32}(\alpha_0\beta_1 - \alpha_1\beta_0)^2.
\end{equation}
So if $\Psi$ is singular, we must have
\begin{equation}\label{eq:alphabeta}
	\alpha_0\beta_1 = 2\alpha_1\beta_0
\end{equation}
up to an interchange of $\alpha$ and $\beta$.  The algebraic condition 
\eqref{eq:alphabeta} corresponds to two real equations, so it can be 
expected to hold on a subset of codimension two in $M$ for generic 
spacetimes.  In particular, the solution set has empty interior.

\providecommand{\bysame}{\leavevmode\hbox to3em{\hrulefill}\thinspace}

\end{document}